\documentclass[preprint,amsmath,amssymb,aps,showpacs]{revtex4}

\newcommand{\UQ}{ARC Centre of Excellence for Quantum-Atom Optics, 
School of Mathematics and Physics, University of Queensland, QLD 4072, Australia.}
\newcommand{\NZ}{Jack Dodd Centre for Quantum Technology, Department of Physics, University of Otago, P. O. Box 56, Dunedin, New Zealand}
\newcommand{\VIR}{Department of Physics, University of Virginia, 382 McCormick Road, Charlottesville, Virginia 22904-4714, USA}
\usepackage{bm}
\usepackage{graphicx}
\usepackage{bm}
\usepackage{color}
\usepackage{amsfonts}
\usepackage{amsbsy}

\newcommand{\jpb}{J. Phys. B }

\newcommand{\jpa}{J. Phys. A }

\newcommand{\josas}{J. Opt. Soc. Am.}
\newcommand{\njp}{New J. Phys.}

\begin{document}
\title{Analysis of a continuous-variable quadripartite cluster state from a single optical parametric oscillator}

\author{S.~L.~W. Midgley}
\affiliation{\UQ}
\author{A.~S. Bradley}
\affiliation{\NZ}
\author{O.~Pfister}
\affiliation{\VIR}
\author{M.~K. Olsen}
\affiliation{\UQ}

\date{\today}

\begin{abstract}

We examine the feasibility of generating continuous-variable multipartite entanglement in an intra-cavity quadruply concurrent downconversion scheme that has been proposed for the generation of cluster states by Menicucci \textit{et al.} [Physical Review Letters \textbf{101}, 130501 (2008)]. By calculating optimized versions of the van Loock-Furusawa correlations we demonstrate genuine quadripartite entanglement and investigate the degree of entanglement present. Above the oscillation threshold the basic cluster state geometry under consideration suffers from phase diffusion. We alleviate this problem by incorporating a small injected signal into our analysis. Finally, we investigate squeezed joint operators. While the squeezed joint operators approach zero in the undepleted regime, we find that this is not the case when we consider the full interaction Hamiltonian and the presence of a cavity. In fact, we find that the decay of these operators is minimal in a cavity, and even depletion alone inhibits cluster state formation.  \end{abstract}


\maketitle

\section{Introduction}

Cluster states are a class of graph states \cite{graph} which are of central importance as a resource state for use in one-way, or measurement-based, quantum computing \cite{Jozsa_review}. This type of quantum computation, proposed in 2001 by Raussendorf and Briegel \cite{raussen_briegel_1,raussen_briegel_3,raussen_browne_briegel}, differs significantly from the traditional circuit model of quantum computing in which unitary evolution is achieved via a sequence of operations on single qubits \cite{deutsch, barenco}. In order to realize one-way quantum computing, a cluster state is generated and then a sequence of measurements are performed on this highly entangled multipartite state \cite{nielsen}. 

Most quantum computing proposals are based on qubits. Experiments have also been performed with qubits. In particular, Grover's algorithm has been implemented using an optical one-way quantum computer \cite{zeilinger}. This approach relied on generating a four-qubit cluster state using a number of independent optical parametric oscillators (OPOs) and beam splitters, and then performing measurements on this state. 

Lloyd and Braunstein \cite{lloyd} were the first to highlight the potential use of continuous-variables (CV) in quantum information. Multipartite CV entanglement has also been extensively studied \cite{loock2000,loock2003,aoki,jing}. Since then, with the development of one-way quantum computing, the notion of using CV cluster states as a potential resource has arisen. Proposals specific to CV cluster state quantum computing and the generation of CV cluster states are numerous \cite{nielsen,men3,loock,jzhang06,gu2007,mile_cluster_states,men2}. They include schemes based on using a combination of single-mode squeezers and quantum non-demolition (QND) gates \cite{men3} or schemes that rely on single-mode squeezers and a network of beam splitters \cite{gu2007}. This field continues to attract interest and recently another proposal, based on realizing a CV cluster state using only a single QND gate, was put forward \cite{men2010}. Experimental efforts to generate CV cluster states have also taken place, with the first quadripartite cluster state generated in 2007 by Su \textit{et al.} \cite{Kpeng2007}. Similar experiments have been performed by Yukawa \textit{et al.} \cite{furu2008}.

Recent efforts \cite{zaidi,men2} have also focussed on the possibility of using a single OPO as a means of generating CV multipartite entanglement, and in turn, a CV cluster state. In this scheme, a single optical cavity is pumped by a number of field modes. The different modes of the resonator represent the multipartite entangled systems in the scheme. According to \cite{men2}, the method generates a cluster state with the quadratures of the optical frequency comb of the OPO acting as a quantum computer register. In this article we extend this analysis and consider the feasibility of such a scheme. 

In \cite{zaidi} a single multimode OPA pumped by two field modes is considered in the undepleted pump approximation. A correspondence is shown between the CV multipartite entangled output from this scheme and a CV square-cluster state. It is the square-cluster OPO scheme proposed in \cite{zaidi,men2} that we consider in our work. Qubit graph states analogous to the CV square-cluster state have been studied extensively \cite{hein}. However, apart from \cite{men1} no study of the entanglement properties of the square-cluster OPO scheme has been conducted. Specifically, we investigate this concurrent system in order to verify the presence of CV quadripartite entanglement and determine whether or not a CV cluster state is produced. This builds on our previous analysis of a similar scheme also based on concurrent nonlinearities \cite{midgley}, in which we showed that quadripartite entanglement is present for the case of an OPO pumped by four field modes. Furthermore, our work here extends tripartite schemes proposed in Refs.~\cite{ash,pooser,xie} by utilizing quadruply concurrent nonlinearities. 

This paper is organized as follows. Section II provides an overview of the defining relation for CV cluster states, presents the Hamiltonian for the scheme under consideration and describes the connection between this scheme and the generation of a cluster state. In addition, Sec.~\ref{III} describes the van Loock-Furusawa (VLF) criteria which are used as a means of quantifying quadripartite entanglement. Section III considers the interaction Hamiltonian in the undepleted pump approximation and gives the VLF correlations under this approximation. In Sec. IV we present the full equations of motion for the system and calculate the VLF correlations without a cavity present, using the positive-$P$ method. We also find the field intensities and compare these to the intensities in the undepleted approximation. Section V provides an overview of the linearized fluctuation analysis used in this work to calculate the measurable output fluctuation spectra from the cavity. These output spectra are also found in Sec. V and used to demonstrate violation of the optimized van Loock-Furusawa criteria and hence, demonstrate quadripartite entanglement. The spectra are obtained above and below the oscillation threshold and the steady-state solutions above and below the  threshold are also found along with an expression for the critical pumping. Finally, in Sec. VI we consider whether or not the entangled output beams produced in the proposed scheme do in fact constitute a cluster state, by using the defining relation for cluster states and the squeezed joint quadrature operators.

\section{Generation of a square-cluster state from an OPO}
\label{scheme}

A CV multi-mode entangled state can be classified as a cluster state if the defining relation presented in Refs. \cite{gu2007,men1} is satisfied. To consider this definition, we first define quadrature field operators for each mode as,

\begin{equation}
\hat{X}_{i}=\hat{a}_{i}+\hat{a}^{\dagger}_{i}, \hspace{0.2cm} \hat{Y}_{i}=-i(\hat{a}_{i}-\hat{a}^{\dagger}_{i})
\label{eqn1}
\end{equation}

\noindent such that $[\hat{X}_{i},\hat{Y}_{i}]=2i$. We also let $\boldsymbol{X}$ and $\boldsymbol{Y}$ represent column vectors of the amplitude and phase quadratures, respectively, for each field mode. The definition of a CV cluster state is then any Gaussian state whose quadratures satisfy

\begin{equation}
\boldsymbol{Y} - A\boldsymbol{X} \longrightarrow \boldsymbol{0},
\label{eqn2}
\end{equation}

\noindent where $A$ is the adjacency matrix representing the graph of a given CV state and the arrow specifies that the condition holds in the limit of infinite (or large) squeezing. When this condition is satisfied for a particular $A$-matrix, the CV state is a cluster state. The adjacency matrix can be weighted and represents the couplings between different nodes on the graph representing the cluster state.

\subsection{Physical Description and Hamiltonian}
\label{subsection}

\begin{figure}[tbhp]
\begin{center}
\includegraphics[width=0.5\textwidth]{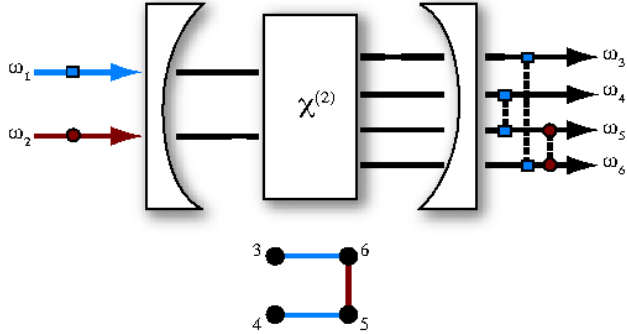}
\end{center}
\caption{(Color online) A $\chi^{(2)}$ crystal inside a pumped Fabry-P\'{e}rot cavity. Pump lasers drive two intracavity modes with frequencies $\omega_{1}$ and $\omega_{2}$ (represented by circles and squares), which are down-converted to four output modes with frequencies $\omega_{3}$, $\omega_{4}$, $\omega_{5}$ and $\omega_{6}$.}
\label{fig1}
\end{figure}

The system we model in this paper is comprised of an optical cavity containing a $\chi^{(2)}$ non-linear crystal. The optical cavity is pumped by two field modes to produce four low-frequency entangled output modes at frequencies $\omega_{3},\omega_{4},\omega_{5},\omega_{6}$. Mode 1 is pumped at a particular frequency and polarization such that it produces modes 3 and 6, as well as modes 4 and 5. Mode 2 is pumped such that it gives rise to modes 5 and 6. A schematic of the setup is shown in Fig.~\ref{fig1}. 

The Hamiltonian for this six-mode system is given by

\begin{equation}
{\cal H} ={\cal H}_{int}+{\cal H}_{free}+ {\cal H}_{pump}+{\cal H}_{bath},
\label{eq:Hfull}
\end{equation}

\noindent where the interaction Hamiltonian is
\begin{eqnarray}
{\cal H}_{int} = i\hbar[\chi_{1}\hat{a}_{1}\hat{a}_{4}^{\dagger}\hat{a}_{5}^{\dagger} + \chi_{1}\hat{a}_{1}\hat{a}_{3}^{\dagger}\hat{a}_{6}^{\dagger} +\chi_{2}\hat{a}_{2}\hat{a}_{5}^{\dagger}\hat{a}_{6}^{\dagger}]+ \textnormal{h.c.},
\label{Hint}
\end{eqnarray}
with the $\chi_{i}$ representing the effective nonlinearities and $\hat{a}_{i}$ and $\hat{a}^{\dagger}_{i}$ denoting the bosonic annihilation and creation operators, respectively, for the intra-cavity modes at frequencies $\omega_{i}$. The pumping Hamiltonian which describes the cavity driving fields, in the appropriate rotating frame is

\begin{equation}
{\cal H}_{pump} = i\hbar\sum_{i=1}^{2}\left[\epsilon_{i}\hat{a}_{i}^{\dag}-\epsilon_{i}^{\ast}\hat{a}_{i}\right],
\label{eq:Hpump}
\end{equation}

\noindent and the cavity damping Hamiltonian is given by

\begin{equation}
{\cal H}_{bath}= \hbar\sum_{i=1}^{6}\left[\hat{\Gamma}_{i}\hat{a}_{i}^{\dag}+\hat{\Gamma}_{i}^{\dag}\hat{a}_{i}\right],
\label{eq:Hres}
\end{equation}

\noindent where $\epsilon_{i}$ are the classical pumping laser amplitudes for modes $i$, and the $\hat{\Gamma}_{i}$ are the annihilation operators for bath quanta, representing losses through the cavity mirrors. 

\subsection{The Undepleted Pump Approximation}

Prior to studying the full Hamiltonian in the presence of an optical cavity, it is useful to consider the properties of the Hamiltonian within the undepleted pump approximation. This approximation assumes that all the high frequency pump modes remain highly populated throughout the interaction process, with no depletion taking place. Specifically, with the cavity absent we set $\xi_{1}=\chi_{1}\langle \hat{a}_{1}(0)\rangle$ and $\xi_{2}=\chi_{2}\langle \hat{a}_{2}(0)\rangle$ where $\xi_{i}$ are positive, real constants. Under this approximation, the interaction Hamiltonian can be written as follows,

\begin{equation}
{\cal H}_{int}= i\hbar \xi \sum_{j=m,n}G_{mn}\left[\hat{a}_{m}^{\dagger}\hat{a}_{n}^{\dagger} -\hat{a}_{m}\hat{a}_{n}\right],
\label{Halternative}
\end{equation}

\noindent where we assume $\xi=\xi_{i} (i=1,2)$, $j$ represents all permutations of the low frequency modes and $G_{mn}$ are the components of the matrix,

	\begin{equation} G=\left[ \begin{array} {c c c c} 
	0&0&0&1\\
	0&0&1&0\\
	0&1&0&1\\
	1&0&1&0\end{array}\right].
	\label{Gmn}
	\end{equation}

Inspecting the form of this $G$-matrix for the system, it can be seen that it corresponds to the graph in Fig.~\ref{fig1}. The four nodes represent the four low frequency modes and the lines connecting the nodes represent the nonlinear coupling of the modes. Such a graph, representing the Hamiltonian of Eq.~(\ref{Halternative}), has been shown to be equivalent to a four node square-cluster state \cite{men1,zaidi}.

\subsection{Criteria for Quadripartite Entanglement}
\label{III}

To determine whether or not a square-cluster state generated by the single OPO scheme presented in Sec.~\ref{scheme} is fully inseparable, it is possible to construct multipartite entanglement witnesses. These are observables that allow one to distinguish multipartite entangled states from separable states. In order to detect CV multipartite entanglement we use the set of sufficient conditions proposed by van Loock and Furusawa (VLF) \cite{furu}, which are a generalization of the conditions for CV bipartite entanglement \cite{Duan, Simon}. As shown in \cite{midgley}, these conditions may be optimized for the verification of genuine quadripartite entanglement. It should be noted that other multi-partite entanglement witnesses also exist \cite{eisert2006}. 

Using the quadrature definitions in Eq.~(\ref{eqn1}), the optimized inequalities which must be simultaneously violated by the low frequency modes in order to demonstrate CV quadripartite entanglement are given by,

\begin{eqnarray}
\label{V36}
V(\hat{X}_{3}-\hat{X}_{6}) + V(\hat{Y}_{3} + g_{4}\hat{Y}_{4} + g_{5}\hat{Y}_{5} + \hat{Y}_{6}) \ge 4,\\
\label{V45}
V(\hat{X}_{4}-\hat{X}_{5}) + V(g_{3}\hat{Y}_{3} + \hat{Y}_{4} + \hat{Y}_{5} + g_{6}\hat{Y}_{6}) \ge 4,\\
\label{V56}
V(\hat{X}_{5}-\hat{X}_{6}) + V(g_{3}\hat{Y}_{3} + g_{4}\hat{Y}_{4} + \hat{Y}_{5} + \hat{Y}_{6}) \ge 4,\\\nonumber
\end{eqnarray}

\noindent where $V(\hat{A})=\langle\hat{A}^{2}\rangle-\langle\hat{A}\rangle^{2}$ denotes the variance and the $g_{i}  (i=3,4,5,6)$ are arbitrary real parameters that are used to optimize the violation of these inequalities. Eq.~(\ref{V36}) and Eq.~(\ref{V45}) are minimized with respect to $g_{4,5}$ and $g_{3,6}$, respectively. We then solve the resulting equations to obtain the optimized expressions,

\begin{eqnarray}
g_{3} &=& \frac{V_{6}(V_{34}+V_{35})-V_{36}(V_{46}+V_{56})}{V^{2}_{36}-V_{3}V_{6}},\\
g_{4} &=& \frac{V_{5}(V_{34}+V_{46})-V_{45}(V_{35}+V_{56})}{V^{2}_{45}-V_{4}V_{5}},\\
g_{5} &=& \frac{V_{4}(V_{35}+V_{56})-V_{45}(V_{34}+V_{46})}{V^{2}_{45}-V_{4}V_{5}},\\
g_{6} &=& \frac{V_{3}(V_{46}+V_{56})-V_{36}(V_{34}+V_{35})}{V^{2}_{36}-V_{3}V_{6}},
\end{eqnarray}

\noindent where 

\begin{equation}
V_{ij}=\frac{\langle \hat{Y}_{i}\hat{Y}_{j}\rangle+\langle \hat{Y}_{j}\hat{Y}_{i}\rangle}{2}-\langle \hat{Y}_{i}\rangle\langle \hat{Y}_{j}\rangle
\end{equation} 

\noindent represents the covariances. For the case where $i=j$ the covariance, denoted $V_{i}$, reduces to the usual variance, $V(\hat{Y}_{i})$. 

\section{The Heisenberg Equations}

Within the undepleted pump approximation we can calculate the VLF criteria from Sec.~\ref{III} in order to verify the presence of multipartite entanglement.
The Heisenberg equations of motion for the field operators are given by,

\begin{eqnarray}
\frac{d\hat{a}_{3}}{dt}&=&\xi_{1}\hat{a}_{6}^{\dagger},\\
\frac{d\hat{a}_{4}}{dt}&=&\xi_{1}\hat{a}_{5}^{\dagger},\\
\frac{d\hat{a}_{5}}{dt}&=&\xi_{1}\hat{a}_{4}^{\dagger} + \xi_{2}\hat{a}_{6}^{\dagger},\\
\frac{d\hat{a}_{6}}{dt}&=&\xi_{1}\hat{a}_{3}^{\dagger} + \xi_{2}\hat{a}_{5}^{\dagger},
\end{eqnarray}

\noindent and in turn these equations can be written in terms of the quadrature operators,

\begin{eqnarray}
\frac{d\hat{X}_{3}}{dt}&=&\xi_{1}\hat{X}_{6},\\
\frac{d\hat{Y}_{3}}{dt}&=&-\xi_{1}\hat{Y}_{6},\\
\frac{d\hat{X}_{4}}{dt}&=&\xi_{1}\hat{X}_{5},\\
\frac{d\hat{Y}_{4}}{dt}&=&-\xi_{1}\hat{Y}_{5},\\
\frac{d\hat{X}_{5}}{dt}&=&\xi_{1}\hat{X}_{4}+\xi_{2}\hat{X}_{6},\\
\frac{d\hat{Y}_{5}}{dt}&=&-\xi_{1}\hat{Y}_{4}-\xi_{2}\hat{Y}_{6},\\
\frac{d\hat{X}_{6}}{dt}&=&\xi_{1}\hat{X}_{3}+\xi_{2}\hat{X}_{5},\\
\frac{d\hat{Y}_{6}}{dt}&=&-\xi_{1}\hat{Y}_{3}-\xi_{2}\hat{Y}_{5}.
\end{eqnarray}

\noindent We solve these equations to find analytic solutions for the quadrature operators as functions of their initial values, and in turn the optimized VLF criteria can be calculated. Not a great deal is learnt from the exact form of these rather complicated analytic expressions. Hence, we plot the solutions for the optimized VLF criteria for the cases of equal and unequal values of $\xi_{i}$.

\begin{figure}[tbhp]
\includegraphics[width=0.5\textwidth]{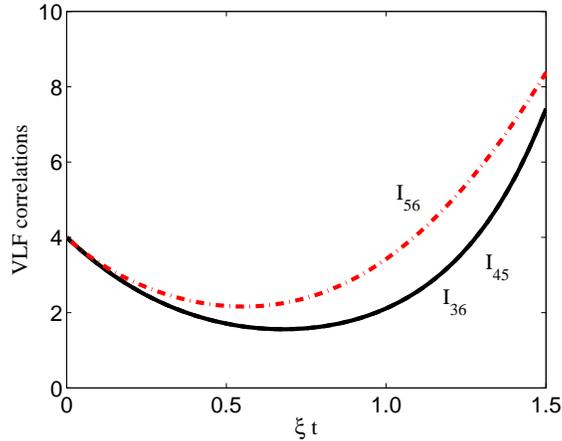}
\caption{(Color online) Optimized van Loock-Furusawa correlations, $\textnormal{I}_{36}$ and $\textnormal{I}_{45}$ (black solid line) and $\textnormal{I}_{56}$ (red dashed-dot line), found by solving the Heisenberg
equations of motion in the undepleted pump approximation.
Having all three of the correlations drop below 4 is sufficient to demonstrate quadripartite entanglement.
All quantities depicted here and in subsequent graphs are dimensionless.}
\label{fig02}
\end{figure}

We first investigate solutions with both $\xi_{i}$ equal. That is, we assume that $\xi=\xi_{1,2}$ and plot the optimized VLF criteria. The correlations $\textnormal{I}_{36}, \textnormal{I}_{45}$ and $\textnormal{I}_{56}$ correspond to Eq.~(\ref{V36}) - Eq.~(\ref{V56}), respectively. Therefore, a value less than four violates the VLF inequalities. In Fig.~\ref{fig02} we see that quadripartite entanglement is present since all three inequalities are simultaneously violated.

\begin{figure}[tbhp]
\includegraphics[width=0.5\textwidth]{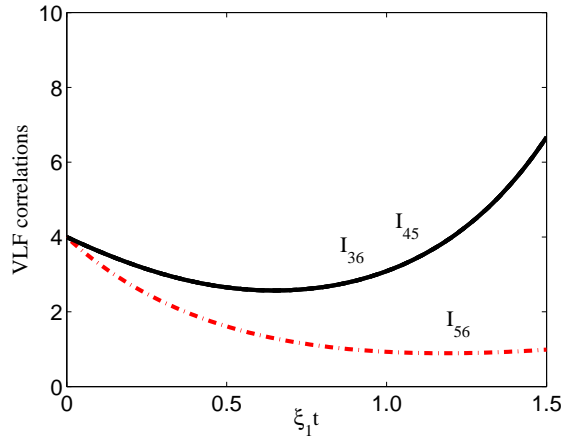}
\caption{(Color online) Optimized van Loock-Furusawa correlations, $\textnormal{I}_{36}$ and $\textnormal{I}_{45}$ (black solid line) and $\textnormal{I}_{56}$ (red dashed-dot line) with $\xi_{2}=0.5\xi_{1}$. }
\label{fig03}
\end{figure}

We also consider solutions with unequal $\xi_{i}$, for the case where $\xi_{2}=0.5\xi_{1}$. The VLF correlations are shown in Fig.~\ref{fig03}. The violation of $\textnormal{I}_{36}$ and  $\textnormal{I}_{45}$ in this case is less than in the symmetric case shown in Fig. \ref{fig02}, however, we still observe a substantial violation of these VLF inequalities. Furthermore, the violation of $\textnormal{I}_{56}$ is greater here than in the symmetric case.

\section{The positive-$P$ equations}

We now turn to an analysis of the setup introduced in Sec.~\ref{subsection} by considering the full interaction Hamiltonian and introducing a pumped, resonant optical cavity to house the nonlinear media. The master equation for this system can be derived using a standard approach \cite{Danbook} and is given by,

\begin{equation}
\frac{\partial \hat{\rho}}{\partial t} = -\frac{i}{\hbar}\Big[\hat{H}_{pump}+\hat{H}_{int}, \hat{\rho}\Big]+\sum_{i=1}^{8}\gamma_{i} {\cal{D}}_{i}[\hat{\rho}]
\label{master}
\end{equation}

\noindent where $\gamma_{i}$ represent the cavity losses at each frequency and $\cal{D}$$_{i}[\hat{\rho}]=2\hat{a}_{i} \hat{\rho}\hat{a}^{\dagger}_{i} -\hat{a}^{\dagger}_{i} \hat{a}_{i} \hat{\rho}-\hat{\rho}\hat{a}^{\dagger}_{i} \hat{a}_{i} $ is the Lindblad superoperator \cite{Danbook} under the zero-temperature Markov approximation. From the master equation it is possible to derive a set of stochastic differential equations (SDEs) and then investigate the intra-cavity dynamics.

We use the positive-$P$ representation \cite{plusP,GardinerQN} to perform a full quantum analysis. In this approach the master equation, Eq.~(\ref{master}), is mapped onto a Fokker-Planck equation (FPE) for the positive-$P$ function to arrive at a set of $c$-number SDEs. To obtain the SDEs it is necessary that the diffusion matrix of the FPE is positive-definite. In the positive-$P$ approach this is achieved by defining two independent stochastic fields $\alpha_{i}$ and $\alpha^{+}_{i}$ and making a correspondence between these operators and the mode operators $\hat{a}_{i}$ and $\hat{a}^{\dagger}_{i}$, respectively. This approach allows us to perform stochastic calculations of normally-ordered operator moments, for example, in the limit of a large number of trajectories $\overline{(\alpha_{j}^{+})^{m}\alpha_{i}^{n}}=\langle:(\hat{a}^{\dagger}_{j})^{m}\hat{a}_{i}^{n}:\rangle$. Therefore, despite being probabilistic, the positive-$P$ method allows for a full quantum treatment of the system when a sufficiently large number of trajectories is used. Following this approach, the resulting 12 $\times$ 12 diffusion matrix is of the form,

\begin{equation}
\boldsymbol{D}=\bordermatrix{&\cr &\bf{0_{4\times4}}&\bf{0_{4\times8}}\cr&\bf{0_{8\times4}}&\boldsymbol{d}},
\end{equation}

\noindent where $\textbf{0}_{r \times c}$ are null matrices and $\boldsymbol{d}$ is an $8\times 8$ non-zero block given by,

\begin{widetext}
\begin{equation}
\boldsymbol{d}=\bordermatrix{&\cr &0&0 &0&0 &0&0 &\chi_{1}\alpha_{1}&0\cr&0&0 &0&0 &0&0 &0&\chi_{1}\alpha_{1}^{+} \cr&0 &0 &0&0 &\chi_{1}\alpha_{1}&0 &0&0\cr&0&0 &0&0 &0&\chi_{1}\alpha_{1}^{+} &0&0\cr&0&0 &\chi_{1}\alpha_{1}&0 &0&0 &\chi_{2}\alpha_{2}&0\cr&0&0 &0&\chi_{1}\alpha_{1}^{+}  &0&0 &0&\chi_{2}\alpha_{2}^{+} \cr&\chi_{1}\alpha_{1}&0 &0&0 &\chi_{2}\alpha_{2}&0 &0&0\cr&0&\chi_{1}\alpha_{1}^{+} &0&0 &0&\chi_{2}\alpha_{2}^{+} &0&0}.
\end{equation}
\end{widetext}

\noindent After factorizing $\boldsymbol{D}$ to find the noise terms, we find the set of It\^o SDEs in the positive-$P$ representation. The evolution equations for the high frequency fields are,

\begin{eqnarray}
\frac{d\alpha_{1}}{dt}&=&\epsilon_{1}-\chi_{1}\alpha_{4}\alpha_{5}-\chi_{1}\alpha_{3}\alpha_{6}-\gamma_{1}\alpha_{1},\nonumber\\
\frac{d\alpha_{2}}{dt}&= & \epsilon_{2}-\chi_{2}\alpha_{5}\alpha_{6}-\gamma_{2}\alpha_{2},\nonumber\\
\label{posp1}
\end{eqnarray}

\noindent as well as the equations found by interchanging $\alpha_{i}$ and $\alpha_{i}^{+}$. For the low frequency fields we find,

\begin{eqnarray}
\frac{d\alpha_{3}}{dt}&=& \chi_{1}\alpha_{1}\alpha_{6}^{+}-\gamma_{3}\alpha_{3}+\sqrt{\frac{\chi_{1}\alpha_{1}}{2}}(\eta_{9}(t)+i\eta_{10}(t)),\nonumber\\
\frac{d\alpha_{4}}{dt}&=&\chi_{1}\alpha_{1}\alpha_{5}^{+}-\gamma_{4}\alpha_{4}+\sqrt{\frac{\chi_{1}\alpha_{1}}{2}}(\eta_{5}(t)+i\eta_{6}(t)),\nonumber\\
\frac{d\alpha_{5}}{dt}&=&\chi_{1}\alpha_{1}\alpha_{4}^{+}+\chi_{2}\alpha_{2}\alpha_{6}^{+}-\gamma_{5}\alpha_{5}+\sqrt{\frac{\chi_{1}\alpha_{1}}{2}}(\eta_{5}(t)-i\eta_{6}(t))\nonumber\\&+&\sqrt{\frac{\chi_{2}\alpha_{2}}{2}}(\eta_{1}(t)+i\eta_{2}(t)),\nonumber\\
\frac{d\alpha_{6}}{dt}&=& \chi_{1}\alpha_{1}\alpha_{3}^{+}+\chi_{2}\alpha_{2}\alpha_{5}^{+}-\gamma_{6}\alpha_{6}+\sqrt{\frac{\chi_{1}\alpha_{1}}{2}}(\eta_{9}(t)-i\eta_{10}(t))\nonumber\\&+&\sqrt{\frac{\chi_{2}\alpha_{2}}{2}}(\eta_{1}(t)-i\eta_{2}(t)),\nonumber\\
\label{posp}
\end{eqnarray}

\noindent and also the equations found by swapping $\alpha_{i}$ with $\alpha_{i}^{+}$ and $\eta_{i}(t)$ with $\eta_{i+2}(t)$. The $\eta_{i}(t)$ are real, independent, Gaussian noise terms which have the correlations $\overline{\eta_{i}(t)}=0$ and $\overline{\eta_{i}(t)\eta_{j}(t^{\prime})}=\delta_{ij}\delta(t-t^{\prime})$. We assume throughout this article that all the intracavity modes are resonant with the cavity and as a result no detuning terms are included.

\begin{figure}[tbhp]
\includegraphics[width=0.5\textwidth]{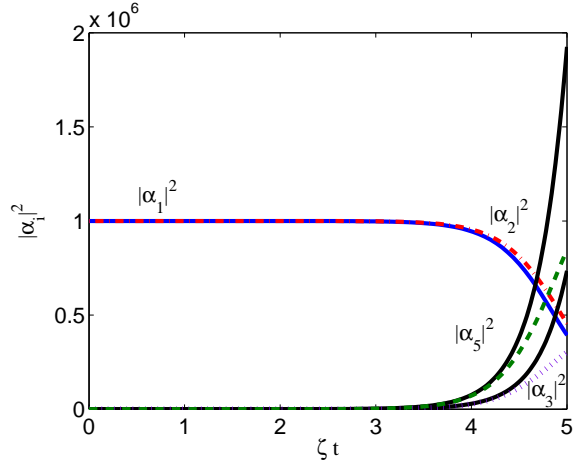}
\caption{(Color online) Intensities of the high (red dashed-dot and blue solid lines) and low frequency modes (dashed green and small dashed purple lines) calculated using the positive-$P$ equations without a cavity present. The number of trajectories is 300,000. The parameters used are $\chi_{1,2}=0.01$, $\alpha_{1,2}(0)=1 \times 10^{3}$ and $\alpha_{3,4,5,6}(0)=0$. The black solid lines show the low frequency modes in the undepleted pump approximation. }
\label{fig04}
\end{figure}

An initial insight into the downconversion processes can be garnered by neglecting the pump and loss terms in Eqs.~(\ref{posp1}) and Eqs.~(\ref{posp}) momentarily and simply looking at the dynamics with depletion present. The results of such a positive-$P$ simulation are shown in Fig.~\ref{fig04} and compared to the undepleted pump approximation results for the low frequency modes. Specifically, we plot the intensities of the fields where the horizontal axis is a scaled interaction time, with $\zeta=\chi |\alpha_{1,2}(0)|$ and $\chi=\chi_{i} (i=1,2)$. As expected, the undepleted pump results deviate from the positive-$P$ results as depletion becomes significant. 

We also calculate the VLF correlations using the positive-$P$ equations without a cavity present. In particular, we use 300,000 trajectories and all other parameters are the same as in Fig.~\ref{fig04}. The results compare well with the undepleted pump results shown in Fig.~\ref{fig02} and are not visibly different for a large number of trajectories.

\section{Linearized Fluctuation Analysis}

Returning to our analysis of the more experimentally relevant system introduced in Sec.~\ref{subsection}, we can consider the full quantum dynamics with depletion present and where the interaction occurs inside an optical cavity. This type of scheme is currently under experimental study by one of us \cite{men1,zaidi} at the University of Virginia. 

We undertake a linearized fluctuation analysis \cite{Danbook} to obtain output spectral correlations for the cavity from the intracavity spectra. This is achieved by first linearizing the Eqs.~(\ref{posp1}) and~(\ref{posp}) around the classical steady-state solutions. In the usual manner \cite{Danbook,SMCrispin}, we then find a set of evolution equations for the fluctuations. To begin we neglect the noise terms in Eq.~(\ref{posp}) so that $\alpha_{i}^{+} \rightarrow \alpha_{i}^{*}$. We then set $\alpha_{i}=\bar{\alpha}_{i}+\delta\alpha_{i}$, where $\bar{\alpha}_{i}$ is a mean value and $\delta\alpha_{i}$ represents the fluctuations. This gives a set of classical equations for the mean values and from these it is possible to obtain steady-state solutions. It also allows one to obtain linearized fluctuation equations from which spectral correlations can be obtained. 

We find that an oscillation threshold is present in our symmetric system. Below this threshold we solve the set of classical equations for the mean steady-state values. We find that the stationary solutions below the threshold value are

\begin{eqnarray}
\bar{\alpha}_{i}&=&\frac{\epsilon_{i}}{\gamma_{i}} \hspace{1cm} \textnormal{for} \hspace{0.1cm}i \in \{1,2\},\nonumber\\
\bar{\alpha}_{i}&=&0 \hspace{1cm} \textnormal{for} \hspace{0.1cm}i \in \{3,4,5,6\}.
\end{eqnarray}

Returning to the linearized fluctuation analysis, to first order in the fluctuations the equations of motion for the fluctuations, $\boldsymbol{\delta\alpha}=[\delta\alpha_{1}, \delta\alpha^{+}_{1},\delta\alpha_{2},\delta\alpha^{+}_{2},\dots,\delta\alpha_{6},\delta\alpha^{+}_{6}]^{T}$, are given by,

\begin{equation}
d\boldsymbol{\delta\alpha}=-\boldsymbol{\bar{A}\delta\alpha} dt + \boldsymbol{\bar{B}} d\boldsymbol{W},
\end{equation}

\noindent where $\bf{\bar B}$ is the noise matrix of Eq.~(\ref{posp}) with the steady-state values inserted, $d\boldsymbol{W}$ is a vector of Wiener increments \cite{GardinerQN} and $\boldsymbol{\bar{A}}$ is the drift matrix with the steady-state values inserted as follows,

\begin{equation}
	 \boldsymbol{\bar{A}}=\bordermatrix{ & \cr & \boldsymbol{A}_{1} & \boldsymbol{A}_{2} \cr & \cr & -(\boldsymbol{A}_{2}^{*})^{T} & \boldsymbol{A}_{3} \cr  },
	\label{sarah}
\end{equation}

\noindent where

\begin{equation}
	 \boldsymbol{A}_{1}=\bordermatrix{ & \cr & \gamma_{1}  & 0 & 0 &0 \cr   &0 &  \gamma_{1}  & 0 & 0\cr   &0 & 0 & \gamma_{2}  & 0  \cr   &0  & 0 & 0 & \gamma_{2} \cr },
\end{equation}

\begin{widetext}
\begin{equation}
	 \boldsymbol{A}_{2}=\bordermatrix{ & \cr & \chi_{1}\bar\alpha_{6} & 0  & \chi_{1}\bar\alpha_{5} & 0  &\chi_{1}\bar\alpha_{4} & 0 & \chi_{1}\bar\alpha_{3}  & 0\cr  & 0  & \chi_{1}\bar\alpha_{6}^{*}  & 0  & \chi_{1}\bar\alpha_{5}^{*} & 0& \chi_{1}\bar\alpha_{4}^{*}  & 0 & \chi_{1}\bar\alpha_{3}^{*} & \cr  & 0 & 0 &  0  & 0  & \chi_{2}\bar\alpha_{6} & 0 &\chi_{2}\bar\alpha_{5}  & 0&\cr   &0  & 0 &0  & 0 & 0 & \chi_{2}\bar\alpha_{6}^{*} &0&\chi_{2}\bar\alpha_{5}^{*}&\cr },
	\label{rr}
\end{equation}

\noindent and

\begin{equation}
	 \boldsymbol{A}_{3}=\bordermatrix{ & \cr & \gamma_{3} & 0  & 0 & 0 & 0 & 0 & 0  & -\chi_{1}\bar\alpha_{1} \cr  & 0  &  \gamma_{3}  & 0  & 0 & 0 & 0&  -\chi_{1}\bar\alpha_{1}^{*} & 0\cr  & 0 & 0  &  \gamma_{4}  & 0  & 0 & -\chi_{1}\bar\alpha_{1} & 0 & 0\cr  & 0  & 0& 0 & \gamma_{4} & -\chi_{1}\bar\alpha_{1}^{*}   &  0 & 0 & 0 & \cr  & 0 & 0  & 0 & -\chi_{1}\bar\alpha_{1}  &  \gamma_{5}& 0 & 0 & -\chi_{2}\bar\alpha_{2}\cr  & 0 & 0& -\chi_{1}\bar\alpha_{1}^{*}   & 0 & 0  &  \gamma_{5} &-\chi_{2}\bar\alpha_{2}^{*} & 0& \cr & 0  & -\chi_{1}\bar\alpha_{1} & 0  & 0 & 0 & -\chi_{2}\bar\alpha_{2} &  \gamma_{6} &0 &\cr& -\chi_{1}\bar\alpha_{1}^{*}  & 0 & 0  & 0 &  -\chi_{2}\bar\alpha_{2}^{*} & 0  & 0&  \gamma_{6}}.
	\label{yy}
\end{equation}

\end{widetext}

For the linearized fluctuation analysis to be valid the fluctuations must remain small compared to the mean values and the eigenvalues of the drift matrix $\boldsymbol{\bar{A}}$ must have no negative real part. The eigenvalues are given by,

\begin{eqnarray}
\lambda_{1,2}&=&\gamma_{1},\nonumber\\
\lambda_{3}&=&\gamma_{2}-\frac{1}{2}\Big[\frac{\epsilon_{2}\chi_{2}}{\gamma_{2}}+\sqrt{\Big(\frac{2\epsilon_{1}\chi_{1}}{\gamma_{1}}\Big)^{2}+\Big(\frac{\epsilon_{2}\chi_{2}}{\gamma_{2}}\Big)^{2}}\Big],\nonumber\\
\lambda_{4}&=&\gamma_{2}+\frac{1}{2}\Big[\frac{\epsilon_{2}\chi_{2}}{\gamma_{2}}-\sqrt{\Big(\frac{2\epsilon_{1}\chi_{1}}{\gamma_{1}}\Big)^{2}+\Big(\frac{\epsilon_{2}\chi_{2}}{\gamma_{2}}\Big)^{2}}\Big],\nonumber\\
\lambda_{5}&=&\gamma_{2}+\frac{1}{2}\Big[-\frac{\epsilon_{2}\chi_{2}}{\gamma_{2}}+\sqrt{\Big(\frac{2\epsilon_{1}\chi_{1}}{\gamma_{1}}\Big)^{2}+\Big(\frac{\epsilon_{2}\chi_{2}}{\gamma_{2}}\Big)^{2}}\Big],\nonumber\\
\lambda_{6}&=&\gamma_{2}+\frac{1}{2}\Big[\frac{\epsilon_{2}\chi_{2}}{\gamma_{2}}+\sqrt{\Big(\frac{2\epsilon_{1}\chi_{1}}{\gamma_{1}}\Big)^{2}+\Big(\frac{\epsilon_{2}\chi_{2}}{\gamma_{2}}\Big)^{2}}\Big].
\label{eig}
\end{eqnarray}

\begin{figure}[tbhp]
\includegraphics[width=0.45\textwidth]{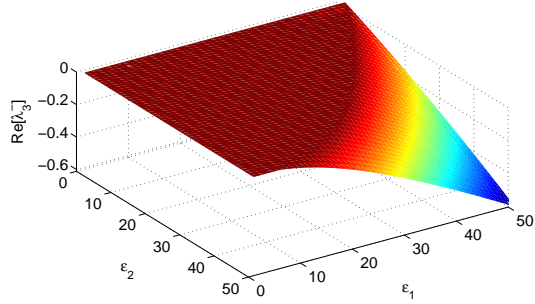}
\caption{(Color online) The region of stability (the plateau) and the transition to instability for a range of pump amplitudes, $\epsilon_{1}$ and $\epsilon_{2}$, found by investigating the behavior of the negative real part of the eigenvalue $\lambda_{3}$, denoted Re[$\lambda_{3}^{-}$].}
\label{fig000}
\end{figure}

\noindent There are six other eigenvalues but each of these is degenerate with one of the eigenvalues in Eq.~(\ref{eig}). From these expressions it is clear that only $\lambda_{3,4}$ can have a negative real part. For our chosen parameters and in the pump range $0 < \epsilon_{1,2} < 100$, it is the eigenvalue $\lambda_{3}$ that has a real part which go from positive to negative. This is depicted in Fig.~\ref{fig000} where we observe a plateau of stability for a range of pump values and the transition to an unstable region, where the negative real part of $\lambda_{3}$ is plotted. In the latter region the linearized fluctuation analysis is not valid. 

In the following we consider the completely symmetric case where the two pumping inputs, $\epsilon_{1,2}$, are equal and given by $\epsilon$, the two nonlinearities, $\chi_{1,2}$, are equal and denoted $\chi$ and finally we assume all the cavity losses are equal and given by $\gamma=\gamma_{i}$ for $i=1,\dots,6$. We can now use the expressions for the eigenvalues, as well as the analytic expressions for the low and high frequency modes below threshold, to find the critical pumping amplitude, $\epsilon_{c}$, at which the oscillation threshold is reached. Furthermore, we confirm via a positive-$P$ simulation that this is the threshold pumping value for which downconversion begins to populate the low frequency modes. For our symmetric system, we find the critical pumping amplitude to be

\begin{equation}
\epsilon_{c}=\frac{\gamma^{2}}{\chi}\Big[ \frac{2}{1+\sqrt{5}}  \Big],
\end{equation}

\noindent where $\epsilon_{c}=61.8$ for our chosen cavity parameters.

If the requirement that the real part of the eigenvalues stay positive is satisfied, the fluctuation equations will describe an Ornstein-Uhlenbeck process \cite{SMCrispin} for which the intracavity spectral correlation matrix is,

\begin{equation}
\boldsymbol{S}(\omega)=(\boldsymbol{\bar{A}}+i\omega \openone)\boldsymbol{\bar{B}\bar{B}}^{T}(\boldsymbol{\bar{A}}^{T}-i\omega \openone)^{-1}.
\label{smatrix}
\end{equation}

\noindent All the correlations required to study the measurable extracavity spectra are contained in this intracavity spectral matrix. Equation~(\ref{smatrix}) is related to the measurable output fluctuation spectra by standard input-output relations for optical cavities \cite{collett}. In particular, the spectral variances and covariances have the general form,

\begin{eqnarray}
\textnormal{S}_{X_{i}}^{out}(\omega)&=&1+2\gamma_{i}S_{X_{i}}(\omega),\nonumber\\
\textnormal{S}_{X_{i},X_{j}}^{out}(\omega)&=&2\sqrt{\gamma_{i}\gamma_{j}}S_{X_{i},X_{j}}(\omega).
\end{eqnarray}

\noindent Similar expressions can be derived for the $\hat{Y}$ quadratures. For brevity, we use $\textnormal{I}^{out}_{ij}$ (ie. any of $\textnormal{I}^{out}_{36}, \textnormal{I}^{out}_{45}, \textnormal{I}^{out}_{56}$) to represent the three output spectral correlations corresponding to the optimized VLF correlations, $\textnormal{I}_{ij}$, of Eq.~(\ref{V36}) to Eq.~(\ref{V56}). That is, the same inequalities as given in Sec.~\ref{III} in terms of variances also hold when expressed in terms of the output spectra. It is these quantities that can be measured in experiments and that we calculate in the remainder of this article.

\subsection{Output spectra below threshold}

\begin{figure}[tbhp]
\includegraphics[width=0.5\textwidth]{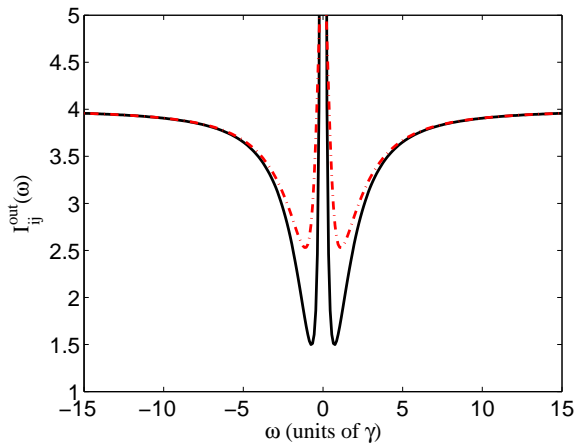}
\caption{(Color online) The spectral correlations below threshold, $\textnormal{I}^{out}_{ij}(\omega)$, for the intracavity quadruply concurrent scheme. The parameter values are $\chi=0.01$, $\gamma=1$ and $\epsilon=0.987\epsilon_{c}$. This value of $\epsilon$ gives best violation of the inequalities.  The correlation $\textnormal{I}^{out}_{56}$ is given by the red dashed-dot line, while $\textnormal{I}^{out}_{36}$ and $\textnormal{I}^{out}_{45}$ are equal and given by the black solid line. }
\label{fig06}
\end{figure}

We now use these steady-state values to calculate the spectra. In Fig.~\ref{fig06} we plot the output spectral correlations, $\textnormal{I}^{out}_{ij}$, as a function of frequency below threshold for a particular pumping rate. The pump rate is chosen to be $\epsilon=0.987\epsilon_{c}$ as this gives the best violation of the inequalities for our choice of parameters. The correlations $\textnormal{I}^{out}_{36}$ and $\textnormal{I}^{out}_{45}$ are equal and give the maximum violation of the inequalities. The correlation $\textnormal{I}^{out}_{56}$ is also shown. We observe that the spectra bifurcates such that no entanglement is present close to zero frequency. For large frequencies $\textnormal{I}^{out}_{ij}(\omega) \rightarrow 4$. This is the uncorrelated limit for the optimized expressions. Collectively, however, these three output spectral correlations confirm that quadripartite entanglement is present below threshold, since all $\textnormal{I}^{out}_{ij}$ drop below four over a range of frequencies.

\subsection{Output spectra above threshold}

It is not as straightforward to obtain analytic expressions for the low and high frequency modes above threshold. As was seen from the graph in Fig.~\ref{fig1}, modes 3 and 4 are only coupled to 6 and 5, respectively, while 5 and 6 are also coupled with each other. This asymmetry results in an undefined overall phase so that, above the oscillation threshold, the low frequency modes suffer phase diffusion, degrading the measurable correlations. We avoid this by injecting a small signal \cite{milburn,mko2} in one of the low frequency modes, thus stabilizing the overall phase. In particular, we find the steady-state solutions numerically using the positive-$P$ equations with an injected signal, $\epsilon_{3}=0.5$, in the relevant evolution equations. The pump is far more intense than the injected signal, with $\epsilon_{3}$ chosen to be approximately one per cent of the pump strength.

\begin{figure}[tbhp]
\includegraphics[width=0.5\textwidth]{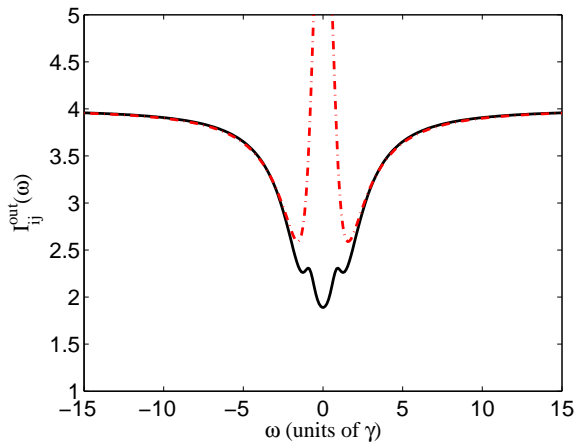}
\caption{(Color online) The spectral correlations above threshold for the intracavity quadruply concurrent scheme. The parameter values are $\chi=0.01$, $\gamma=1$ and $\epsilon=1.49\epsilon_{c}$. The other parameters are the same as in the below threshold case. The correlation $\textnormal{I}^{out}_{56}$ is repesented by the red dashed-dot line. The correlations $\textnormal{I}^{out}_{36}$ and $\textnormal{I}^{out}_{45}$ are equal and represented by the black solid line. }
\label{fig08}
\end{figure}

It is also necessary to make use of an injected signal when calculating the output spectra above threshold. This enables us to calculate the spectra for a range of pump values above threshold. In Fig.~\ref{fig08} we plot the output spectral correlations above threshold for a pumping rate of $\epsilon=1.49\epsilon_{c}$. We see that all three of the VLF inequalities are violated for a range of frequencies and therefore quadripartite entanglement is present above the threshold condition. As in the below threshold case, we also see the spectra bifurcating close to zero frequency and hence not demonstrating entanglement in this region.

\begin{figure}[tbhp]
\includegraphics[width=0.5\textwidth]{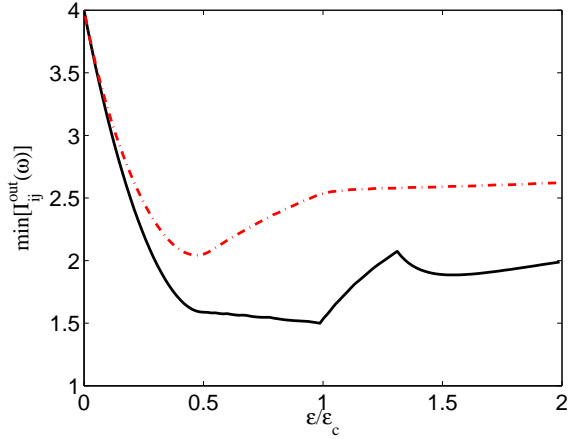}
\caption{(Color online) Maximum quadripartite entanglement as a function of the ratio of the cavity pumping to the pumping threshold. All other cavity parameters are the same as in Fig.~\ref{fig06} and Fig.~\ref{fig08}. Again, the correlations $\textnormal{I}^{out}_{36}$ and $\textnormal{I}^{out}_{45}$ are equal for these parameters and hence cannot be differentiated on the graph. Note that for $\epsilon/\epsilon_{c} = 1$ the validity of the results is limited as the linearized analysis is no longer valid.}
\label{fig09}
\end{figure}

\begin{figure}[tbhp]
\includegraphics[width=0.5\textwidth]{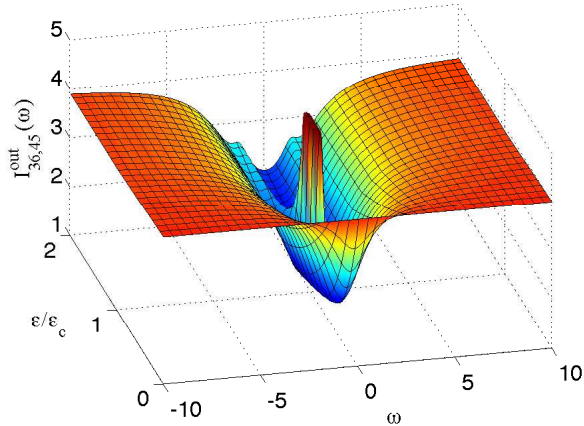}
\caption{(Color online) Plot of the spectral correlations $\textnormal{I}^{out}_{36}$ and $\textnormal{I}^{out}_{45}$ as a function of the pumping rate $\epsilon/\epsilon_{c}$ and frequency $\omega$ (in units of $\gamma$). All other cavity parameters are the same as in Fig.~\ref{fig09}. Note that we crop the peak of the spectra for visualization purposes. }
\label{fig009}
\end{figure}

\begin{figure}[tbhp]
\includegraphics[width=0.5\textwidth]{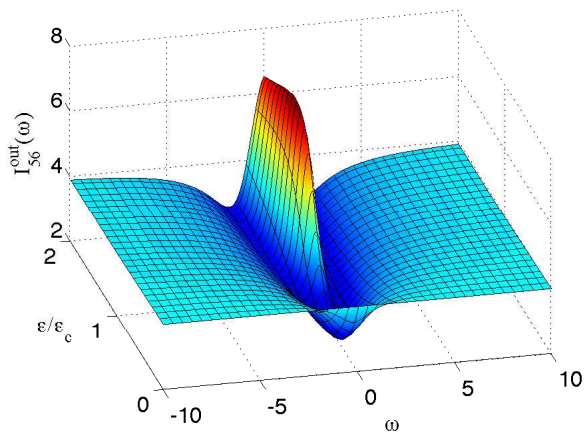}
\caption{(Color online) Plot of the spectral correlation $\textnormal{I}^{out}_{56}$ as a function of the pumping rate $\epsilon/\epsilon_{c}$ and frequency $\omega$ (in units of $\gamma$).  All other cavity parameters are the same as in Fig.~\ref{fig09}.}
\label{fig0009}
\end{figure}

We also calculate the maximum quadripartite entanglement for a range of pump field amplitudes below and above threshold. This result is shown in Fig.~\ref{fig09}, where we plot the minimum value of the output spectra at any frequency, as a function of $\epsilon/\epsilon_{c}$. From this plot it is clear that quadripartite entanglement persists below threshold and well above threshold. In Fig.~\ref{fig009} and Fig.~\ref{fig0009} we plot the VLF correlations as a function of frequency and pumping rate. In Fig.~\ref{fig009}, we crop the peak of the spectra for visualization purposes. The value of this correlation would otherwise increase to a maximum value of $\textnormal{I}^{out}_{36,45} \approx 60$. From these surface plots we can also see the onset of the bifurcation.

\section{Confirming the cluster state defining relation}

Now that we have demonstrated genuine quadripartite entanglement below and above threshold, we turn our attention to whether or not the state generated by the scheme is in fact a cluster state. To do this we employ the defining relation for cluster states, given by Eq.~(\ref{eqn2}). In the undepleted pump regime, there exists a connection between this condition and the squeezed joint operators for the system \cite{zaidi,flamm1,men1,furu2008}. Under certain mode rotations and in the limit of infinite (or large) squeezing the operators for the system give rise to the cluster state equation of Eq.~(\ref{eqn2}) for a square-cluster state. We stress that this equivalence has only been shown to hold within the undepleted pump approximation. We briefly overview the equivalence here. The squeezed joint quadrature operators are simply the eigenvectors of the system found by solving the Heisenberg equations of motion for the system represented by the Hamiltonian in Eq.~(\ref{Halternative}) with the adjacency matrix given by Eq.~(\ref{Gmn}). For our scheme the joint quadrature operators are,

\begin{eqnarray}
O_{1}&=&(-c_{1}\hat{X}_{3}+c_{1}\hat{X}_{4}-\hat{X}_{5}+\hat{X}_{6})e^{-c_{2}r},\nonumber\\
O_{2}&=&(-c_{2}\hat{X}_{3}-c_{2}\hat{X}_{4}+\hat{X}_{5}+\hat{X}_{6})e^{-c_{1}r},\nonumber\\
O_{3}&=&(c_{1}\hat{Y}_{3}+c_{1}\hat{Y}_{4}+\hat{Y}_{5}+\hat{Y}_{6})e^{-c_{2}r},\nonumber\\
O_{4}&=&(c_{2}\hat{Y}_{3}-c_{2}\hat{Y}_{4}-\hat{Y}_{5}+\hat{Y}_{6})e^{-c_{1}r},
\label{o1}
\end{eqnarray}

\noindent where $c_{1}=(\sqrt{5}-1)/2$, $c_{2}=(\sqrt{5}+1)/2$ and $r = \xi t$ is the squeezing parameter. The common eigenstate of these joint operators is a quadripartite entangled state that tends towards a cluster state when $r\rightarrow \infty$. That is, any squeezing operator combinations that are proportional to the squeezing factor $e^{-r}$ will automatically satisfy the cluster state condition of Eq.~(\ref{eqn2}) in the limit of infinite (or large) squeezing. We investigate this by example and compare these operators to the cluster state defining relation, $\boldsymbol{Y} - A\boldsymbol{X} \longrightarrow \boldsymbol{0}$. We choose one possible solution for the adjacency matrix $A$, given by
	\begin{equation} A=\frac{1}{2}\left[ \begin{array} {c c c c} 
	0&0&1&\sqrt{5}\\
	0&0&\sqrt{5}&1\\
	1&\sqrt{5}&0&0\\
	\sqrt{5}&1&0&0\end{array}\right],
	\end{equation}
	
\noindent and which corresponds to a weighted square graph CV cluster state. From Eq.~(\ref{eqn2}) the resulting cluster state equations are,
	
\begin{eqnarray}
Y_{3}-\frac{X_{5}}{2}-\frac{\sqrt{5}X_{6}}{2} \rightarrow 0,\nonumber\\
Y_{4}-\frac{\sqrt{5}X_{5}}{2}-\frac{X_{6}}{2} \rightarrow 0,\nonumber\\
Y_{5}-\frac{X_{3}}{2}-\frac{\sqrt{5}X_{4}}{2} \rightarrow 0,\nonumber\\
Y_{6}-\frac{\sqrt{5}X_{3}}{2}-\frac{X_{4}}{2} \rightarrow 0,
\end{eqnarray}

\noindent where the arrow again represents the limit $r \rightarrow \infty$. In the undepleted regime and in the limit of infinite (or large) squeezing these equations are equivalent  to Eqs.~(\ref{o1}) under certain mode rotations. Specifically, if we rotate modes 5 and 6 by $\pi/2$ by substituting $X\rightarrow Y$ and $Y\rightarrow-X$ the equivalence can be seen \cite{zaidi}. Hence, by simply verifying that the squeezed joint quadrature operators approach zero in the limit $r \rightarrow \infty$, we can determine whether or not the proposed scheme gives rise to a cluster state. As an aside, we actually calculate the variances of the squeezed operators of Eqs.~(\ref{o1}), and ensure that these approach zero, as these are the quantities we have access to in our simulations.  

\begin{figure}[tbhp]
\includegraphics[width=0.5\textwidth]{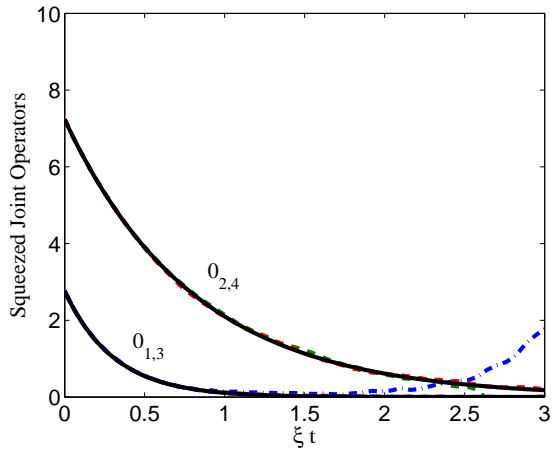}
\caption{(Color online) Squeezed joint operators, $O_{i}$, in the undepleted pump approximation (black solid lines) and positive-$P$ results without a cavity (colored dashed lines). The number of trajectories for the stochastic simulation is 300,000 with $\chi=0.01$ and $\alpha_{1,2}(0)=1 \times 10^{3}$.}
\label{fig10}
\end{figure}

We first confirm that the squeezed operators approach zero as expected \cite{zaidi} in the undepleted case. In Fig.~\ref{fig10} we calculate the squeezed operators from the Heisenberg equations (solid lines). We see that $O_{1}=O_{3}$ and $O_{2}=O_{4}$ and both sets of operators approach zero as $r = \xi t$ approaches infinity. Therefore, the cluster state equation is confirmed in the undepleted regime. In Fig.~\ref{fig10} we also plot the squeezed operators calculated from the positive-$P$ equations in the absence of an optical cavity (dashed lines). We note that the positive-$P$ results go to zero but start to increase from zero at longer times. Moreover, as $O_{2,4}$ approaches zero, $O_{1,3}$ increases and does not approach zero. For the cluster state relation to be satisfied, all squeezing operators must approach zero simultaneously. As seen here this is not the case, except at around $\xi t =2.5$ for the system considered, and hence the required cluster state condition is only partially satisfied once depletion is included. Thus, depletion alone is enough to inhibit cluster state formation. 

\begin{figure}[tbhp]
\includegraphics[width=0.5\textwidth]{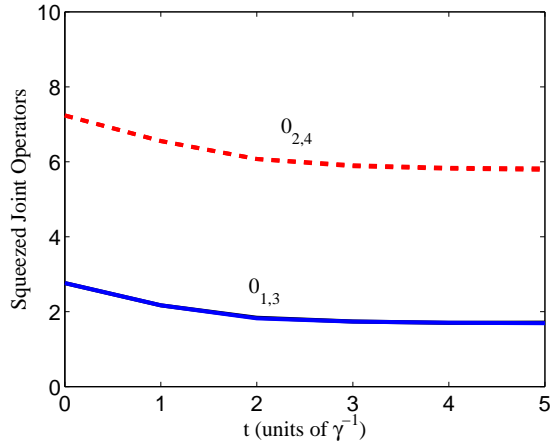}
\caption{(Color online) Squeezed joint operators, $O_{i}$, calculated using the positive-$P$ equations in the presence of a cavity. The operators are calculated for the below threshold case using 50,000 trajectories, with $\epsilon=0.6472\epsilon_{c}$, $\chi=0.01$ and $\gamma=1$.}
\label{fig11}
\end{figure}

In Fig.~\ref{fig11} we plot the squeezing operators for the full interaction Hamiltonian in the presence of a cavity, based on positive-$P$ simulations. This result is more relevant to the scheme that would be realized in the proposed experiment. Unlike the undepleted case, the squeezing operators calculated in this case do not approach zero. Instead, the squeezed operators plateau at non-zero values in the steady-state. We find that the decay of the squeezed joint operators is in fact quite minimal (of order 20\%). This indicates that generating a cluster state from the output of a single OPO may present a challenge.

\section{Conclusions}

We have examined an experimentally feasible quadruply concurrent intracavity scheme as proposed in Refs. \cite{zaidi, men2}. We investigated this system as a potential source of CV quadripartite entanglement and as a candidate for the generation of a CV square-cluster state. We verified the presence of quadripartite entanglement via optimized versions of the well-known VLF correlations. The proposed scheme provided a source of bright entangled output beams above the critical pumping threshold when an injected signal was incorporated into the analysis. Below threshold we also detected quadripartite entanglement with the maximum entanglement predicted near threshold.

We have also calculated the squeezed joint operators to determine if the state produced by the proposed scheme is a CV cluster state. Within the undepleted pump approximation, we confirmed that the squeezed joint operators approached zero in the limit of large squeezing. However, in our analysis of the more experimentally realistic case where depletion is present and a cavity is used to house the nonlinear media, we did not observe that the squeezing operators approached zero as required by the cluster state defining relation. Furthermore, including depletion alone was sufficient to inhibit cluster state formation and once the cavity was also included the decay of the squeezing operators was found to be minimal. This leads us to conclude that the utility of this system as a source of cluster states depends on the degree to which less than perfect squeezing is acceptable. Overall, solution of the undepleted Heisenberg equations of motion can only be used as a general guide to the performance of such a system once it is placed inside an optical cavity. Without the cavity, we also find that an energy-conserving positive-$P$ simulation of the basic downconversion process shows inhibited cluster state formation due to pump depletion.

\section{Acknowledgments}
SLWM and MKO acknowledge the support of the Australian Research Council Centre of Excellence for Quantum-Atom Optics. SLWM would also like to thank the AFGW for their support. ASB is supported by the New Zealand Foundation for Research, Science, and Technology under Contract No.\ UOOX0801. OP is supported by U.S. National Science Foundation grants No.\ PHY-0855632 and No.\ PHY-0555522.


\end{document}